# MODELACIÓN Y VISUALIZACIÓN TRIDIMENSIONAL INTERACTIVA DE VARIABLES ELÉCTRICAS EN CELDAS DE ELECTRO-OBTENCIÓN CON ELECTRODOS BIPOLARES

**César Mena, Ricardo Sánchez, Lautaro Salazar**

*Departamento de Ingeniería Eléctrica*
*Universidad de Concepción*
*Concepción, Chile*

Resumen: El uso de electrodos bipolares flotantes en celdas de electro-obtención de cobre constituye una tecnología no convencional que promete impactos económicos y operacionales. Este artículo presenta una herramienta computacional para la simulación y análisis de tales celdas electroquímicas. Se desarrolla un nuevo modelo para electrodos flotantes y se emplea un método de diferencia finita para obtener la distribución tridimensional de potencial y el campo de densidad de corriente en el interior de la celda. El análisis de los resultados se basa en una técnica para la visualización interactiva de campos vectoriales tridimensionales como líneas de flujo.

Abstract: The use of floating bipolar electrodes in electrowinning cells of copper constitutes a non-conventional technology that promises economic and operational impacts. This paper presents a computational tool for the simulation and analysis of such electrochemical cells. A new model is developed for floating electrodes and a method of finite difference is used to obtain the three-dimensional distribution of the potential and the field of current density inside the cell. The analysis of the results is based on a technique for the interactive visualization of three-dimensional vector fields as streamlines.

Keywords: Modelling, Simulation of Systems and Processes, Visualization, Computer Graphics.

## 1. INTRODUCCIÓN

Debido a los convenientes requerimientos energéticos, la tecnología de electro-obtención (EO) de cobre usando electrodos bipolares promete importantes ventajas económicas y operacionales respecto a la tecnología convencional que usa electrodos unipolares. Sin embargo, para hacer factible un proceso de EO de cobre usando electrodos bipolares, los prototipos existentes deben ser optimizados, ya que producen inaceptables depósitos de cobre, de espesor no homogéneo. La distribución del depósito de cobre esta relacionada directamente con la distribución del campo de densidad de corriente en las superficies catódicas de los electrodos, por lo que es necesario estudiar diferentes estructuras geométricas desde el punto de vista eléctrico. Para tal efecto, es muy útil disponer de una aplicación computacional que permita la simulación y el análisis de celdas de EO, pues se reducen los costos y el tiempo que implican el uso de prototipos, método empleado para el diseño de celdas de EO de cobre basadas en electrodos bipolares.

Este trabajo se inspira en los trabajos previos desarrollados por Bittner *et al*. (1998; 1999), donde se justifica el interés en esta nueva tecnología y se desarrolla un modelo eléctrico para su simulación. En este artículo se presentan los fundamentos de un modelo eléctrico más preciso y elementos de visualización, con los cuales se desarrolló una herramienta para la simulación y análisis interactivo de celdas de EO. La aplicación permite estudiar geometrías rectangulares arbitrarías de celdas que incluyan electrodos unipolares y bipolares flotantes.

## 2. MODELACIÓN ELÉCTRICA DE UNA CELDA DE ELECTRO-OBTENCIÓN

### 2.1 *Fundamentos teóricos del proceso de electrodo*

El proceso electroquímico en el que se basa la EO de cobre extrae el cobre de una solución ácida que contiene sulfato de cobre (electrolito). Cuando se hace circular una corriente DC desde ánodo a cátodo a través del electrolito, el agua se descompone en el ánodo, generando oxígeno gaseoso, y el ión cúprico se reduce, depositándose cobre metálico en el cátodo. Se debe mantener la composición del electrolito aproximadamente constante, mediante un circuito hidráulico que renueve el electrolito.

Para que se produzca el proceso electroquímico en el sentido deseado, se debe aplicar una diferencia de potencial entre los electrodos, la que debe ser mayor que la diferencia entre los potenciales de equilibrio (potenciales de electrodo) de las semireacciones de reducción y oxidación. Si se aplica un voltaje inferior, la reacción ocurre en el sentido opuesto. La forma aproximada de la característica corriente-voltaje de la reacción se muestra en la Fig. 1(a), donde $j$ [A/m$^2$] es la densidad de corriente y $E_R$ es el potencial de equilibrio en condiciones no estándar.

El origen del potencial de electrodo $E_R$, asociado al cambio de fase, se relaciona con la formación de la denominada *doble capa electroquímica* en la interfase metal-electrolito. Cuando se pone en contacto dos fases con conductividades eléctricas significativas se produce una redistribución de carga. La región que contiene esta distribución tiene una extensión que corresponde al radio de los iónes y se conoce como doble capa electroquímica. Esta distribución de carga causa un aumento gradual de potencial en el interior de la doble capa, como se muestra en la Fig. 1(b).

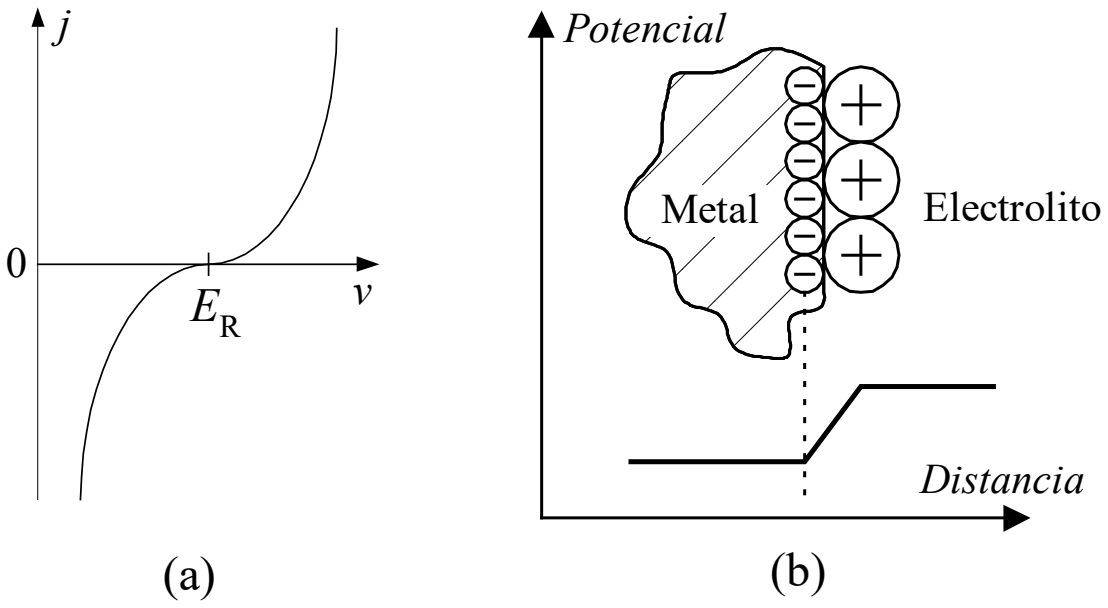

Fig. 1. (a) Característica corriente-voltaje de un proceso de electrodo. (b) Doble capa electroquímica y potencial de electrodo asociado.

### 2.2 *Base de la modelación eléctrica*

La modelación integral de este proceso reviste una gran dificultad, debido a la interacción de variables eléctricas, químicas y mecánicas. Se utiliza como base el modelo desarrollado por Bittner *et al*. (1998; 1999), que como una primera aproximación recurre a la teoría electrostática y considera un campo de corrientes estacionarias, originado por distribuciones de carga constantes. El objetivo es obtener la distribución de la densidad de corriente, en particular en la superficie de los electrodos catódicos, ya que dicha distribución se relaciona directamente con la del depósito de cobre. Se considera que el electrolito es una solución conductora imperfecta homogénea, los electrodos son conductores perfectos y que las paredes de la celda son dieléctricas. Se determina la distribución de potencial eléctrico ($V$) en todo el volumen de la celda. Luego, el vector de campo eléctrico (**E**) y el vector densidad de corriente (**J**) se obtienen respectivamente de:

$$\mathbf{E} = -\nabla V \tag{1}$$

$$\mathbf{J} = \sigma \mathbf{E} \ , \tag{2}$$

donde, $\sigma[\Omega^{-1}m^{-1}]$ es la conductividad eléctrica.

La determinación de $V$ involucra un problema de condiciones de contorno. Se plantean condiciones de contorno en las interfaces donde ocurre cambio de materiales. En el interior de la solución electrolítica, $V$ queda determinado por la ecuación de Laplace:

$$\nabla^2 V = 0 \tag{3}$$

Los electrodos constituyen volúmenes equipotenciales. El dominio de la doble capa electroquímica es no homogéneo. La clave del modelo es considerar esta zona con ancho nulo, definiendo la condición de contorno para la interfase electrodo-electrolito como una discontinuidad de potencial, esto es:

$$V = V_m - DV, \tag{4}$$

donde, $V$ es el potencial en el lado del electrolito y $V_m$ es el *potencial metálico*, el potencial constante del electrodo. La magnitud de la discontinuidad, $DV$, es el potencial de electrodo, medido en el sentido electrodo respecto a la solución electrolítica. La condición de contorno para la interfaz dieléctrico-conductor es que la componente del campo eléctrico normal a la interfase es nula en el lado del conductor, esto es:

$$\nabla V \cdot \hat{\mathbf{n}} = 0, \tag{5}$$

siendo $\hat{\mathbf{n}}$ un vector unitario normal a la superficie de una interfase, orientado hacia el electrolito.

La curva de polarización de la Fig. 1(a) establece la relación entre los potenciales de electrodo y la densidad de corriente en cada punto de la interfaz electrodo-electrolito. Se considera una aproximación lineal de tal relación, la que para ánodo y cátodo se expresa respectivamente como:

$$DV_A = -e_A + K_A \mathbf{J} \cdot \hat{\mathbf{n}} = -e_A - K_A \sigma \nabla V \cdot \hat{\mathbf{n}} \, , \tag{6}$$

$$DV_C = +e_C - K_C \mathbf{J} \cdot \hat{\mathbf{n}} = +e_C + K_C \sigma \nabla V \cdot \hat{\mathbf{n}} \, , \tag{7}$$

donde; $K_A$ y $K_C$ son constantes positivas, $e_A$ y $e_C$ son los potenciales de equilibrio de cada semireaccion.

Si los potenciales metálicos fueran variables independientes, la expresión en diferencias finitas de (3)-(7) en cada punto del dominio genera un sistema de ecuaciones lineales, que, junto a la condición de equipotencialidad, determinaría la distribución de potencial en el interior de la celda. Sin embargo, el potencial metálico de electrodos bipolares flotantes es una variable dependiente. Para resolver el problema electrostático es necesario encontrar esta relación de dependencia e incluirla en el sistema de ecuaciones.

En el modelo desarrollado Bittner *et al*. (1998; 1999) se usan sólo electrodos laminares, en rigor, se debería aplicar la condición de contorno (4) en ambas caras del electrodo. Sin embargo, se usa una solución alternativa que no requiere del potencial metálico. Tal solución fuerza una discontinuidad de potencial entre las dos caras del electrodo bipolar y se consigue manipulando directamente las ecuaciones de diferencia. Los resultados obtenidos con tal modelación concuerdan cualitativamente con algunos resultados experimentales. Sin embargo, la implementación de la condición de contorno usada en el electrodo flotante es contradictoria. En efecto, sólo debería ser usada con potenciales de electrodo constantes en toda la superficie de cada interfase electrodo-electrolito. Si los potenciales de electrodo no son constantes, como en (Bittner *et al*., 1998b; Bittner, 1999), no se satisface la condición (4). Esta contradicción no se hace evidente en los resultados porque las variaciones de los potenciales de electrodos son relativamente pequeñas.

Con objeto de mejorar la precisión de los resultados y, validar el modelo para mejores aproximaciones de la curva de polarización, en la siguiente sección se desarrolla un método para determinar los potenciales metálicos en electrodos flotantes, lo que permite resolver el problema electrostático en forma consistente. Con el modelo mejorado, se pueden implementar electrodos rectangulares con volumen, en los cuales la reacción electroquímica ocurre en toda la superficie del electrodo. Los detalles de la implementación numérica del modelo se encuentran en (Mena, 2000). A diferencia del modelo desarrollado por Bittner *et al*. (1998; 1999), se utiliza un método iterativo para resolver el sistema de ecuaciones. Es la opción más eficiente y precisa para sistemas de ecuaciones con matrices ralas y de gran dimensión, como resulta en este problema. Además, simplifica notablemente la implementación del algoritmo y permite incorporar fácilmente aproximaciones no lineales de la curva de polarización.

### 2.3 *Cálculo del potencial en electrodos flotantes*

La formación de los potenciales de electrodo esta determinada fundamentalmente por los mecanismos de acumulación de carga, debido al flujo de iónes o electrones en la interfase electrodo-electrolito. De la teoría electromagnética se obtiene una relación integral del flujo de electrones que atraviesan una superficie que encierra carga eléctrica (ecuación de Continuidad):

$$\oint_S \mathbf{J}\cdot d\mathbf{s} = -\frac{\partial}{\partial t}\int_{Vol(S)} \rho \, dVol \,, \qquad (8)$$

donde, $d\mathbf{s}$ es un elemento diferencial orientable de superficie que tiene la forma $\hat{\mathbf{n}}\,dS$, siendo $dS$ la magnitud del diferencial de superficie y $\hat{\mathbf{n}}$ el vector normal ya definido. La ecuación relaciona el flujo de densidad de corriente que atraviesa una superficie cerrada con la variación instantánea de carga en el volumen encerrado por la misma. Se considera una superficie cerrada, que encierre un electrodo flotante y que coincida con la interfase electrodo-electrolito por el lado del electrolito, de modo de encerrar completamente la doble capa electroquímica. Se usa la misma referencia para medir los potenciales de electrodo en interfaces anódicas y catódicas, los potenciales de electrodo deben ser calculados apropiadamente en cada región de su interfase con el electrolito (con (6) o (7) ). Luego, el análisis se abstrae del tipo de reacción que ocurre en la superficie y será válido para electrodos unipolares y bipolares.

Como los potenciales de electrodo se definen para el estado de equilibrio dinámico, en el que cesa la acumulación de carga, la relación buscada entre el potencial del electrodo flotante y los potenciales de electrodo se obtiene cuando el término de la derecha en la ecuación (8) es nulo, esto es para

$$\oint_S \mathbf{J}\cdot d\mathbf{s} = \oint_S \sigma\nabla V\cdot d\mathbf{s} = 0 \qquad (9)$$

Los potenciales de electrodo retienen la información relativa al proceso de acumulación de carga. De este modo, al incorporar de alguna forma la condición de contorno establecida en (4) en la ecuación (9), se obtiene la solución buscada. De acuerdo con la teoría del cálculo vectorial (Protter y Morrey, 1964), la cantidad $\nabla V\cdot d\mathbf{s}$ corresponde a una diferencial total. Esta apreciación genera una aproximación numérica de la solución, basándose en el Lema fundamental de derivación. Para $dS$ suficientemente pequeño, el lema permite considerar la siguiente aproximación

$$\nabla V\cdot d\mathbf{s} \approx V(\mathbf{x}+d\mathbf{s}) - V(\mathbf{x}), \text{ cuando } dS \to 0 \qquad (10)$$

Si la superficie $S$ se discretiza en $N$ elementos de superficie $\Delta S$, para $N$ suficientemente grande el integrando de (9) se aproxima usando (10), y al incorporar la condición dada en (4) resulta:

$$\oint_S \mathbf{J}\cdot d\mathbf{s} \approx \sum_{i=1}^{N} \sigma_i \left[ V(\mathbf{x}_i + d\mathbf{s}_i) + DV(\mathbf{x}_i) - V_m \right] = 0 \,, \qquad (11)$$

y en consecuencia:

$$V_m \approx \frac{\sum_{i=1}^{N} \sigma_i \left[ V(\mathbf{x}_i + d\mathbf{s}_i) + DV(\mathbf{x}_i) \right]}{\sum_{i=1}^{N} \sigma_i} \qquad (12)$$

El resultado de (12) es válido en materiales no homogéneos y puede usarse con cualquier tipo de geometría. Las coordenadas $\mathbf{x}+d\mathbf{s}$ corresponden a puntos separados $\Delta S$ unidades de la superficie del electrodo en la dirección $\hat{\mathbf{n}}$. En general, los potenciales $V(\mathbf{x}+d\mathbf{s})$ deben interpolarse.

En el contexto de una solución numérica global del problema electrostático, se puede eliminar la dependencia en los potenciales de electrodos en (12). Si los potenciales se calculan en forma iterativa, en el siguiente orden:

1. Potenciales de electrodo
2. Potenciales del electrolito en la interfase electrodo-electrolito
3. Potenciales del electrolito en el resto del dominio
4. Potenciales de electrodos flotantes,

entonces, al reemplazar apropiadamente (4) en (12), para el caso de electrolito homogéneo, se obtiene la siguiente fórmula recursiva:

$$V_m^{(n)} = V_m^{(n-1)} + \frac{\sum_{i=1}^{N} \left[ V^{(n)}(\mathbf{x}_i + d\mathbf{s}_i) - V^{(n)}(\mathbf{x}_i) \right]}{N}, \quad (13)$$

Se aprecia que $V_m$ converge a un valor determinado cuando la integral de flujo es nula, lo que concuerda con la condición física de equilibrio dinámico.

## 3. VISUALIZACION DE DATOS

Debido a la dimensionalidad de los datos generados por el modelo en estudio, resulta indispensable disponer de herramientas de visualización adecuadas para su análisis. Esta sección describe algunas convenientes herramientas que fueron implementadas en una herramienta de análisis interactivo.

### 3.1 *Visualización de líneas de flujo iluminadas*

La herramienta más importante a desarrollar es una que permita la visualización tridimensional de un campo vectorial estacionario. La representación visual de campos vectoriales es un tema de continua investigación en visualización científica, pero existen técnicas adecuadas para muchos problemas. Para esta aplicación resulta muy conveniente la técnica de visualizar líneas de flujo.

Las líneas de flujo permiten apreciar gráficamente la estructura direccional de un campo vectorial estacionario. Son trayectorias continuas, cuyos vectores tangentes coinciden con el campo vectorial. Se obtienen integrando el campo vectorial a partir de una posición inicial en el espacio (punto semilla). Como el campo resultante esta definido en un dominio discretizado, se debe usar interpolación para calcular el campo en coordenadas intermedias del dominio.

Para obtener una adecuada precisión, se debe usar un método de integración de paso adaptivo con un adecuado control del error. En esta aplicación se usa el método descrito en (Stalling *et al.*, 1995), el cual considera interpolación trilineal y un método de integración Runge-Kutta de cuarto orden.

Comúnmente se enfrentan dos problemas al usan líneas de flujo. Un problema es que no resulta muy obvio como distribuir las líneas de flujo para obtener una adecuada representación del campo vectorial. En problemas electroestáticos se puede lograr una adecuada visualización de la estructura del campo mediante la generación de líneas a partir de superficies equipotenciales. En esta aplicación, los electrodos constituyen adecuadas superficies equipotenciales, en particular porque interesa analizar el campo normal en sus superficies. En consecuencia, se desarrolla un método automático que genera líneas de flujo sobre la superficie de cada electrodo. Se distribuyen homogéneamente puntos semilla en la superficie del electrodo mediante un parámetro de densidad de líneas, ajustable por el usuario. Las líneas se integran desde los puntos semilla hacia el exterior del electrodo y son truncadas por el método de integración adaptivo o cuando se encuentra una singularidad del campo.

El segundo problema se refiere a la limitación del hardware gráfico convencional a la iluminación de primitivas de superficie. La iluminación de las líneas es indispensable para lograr una adecuada percepción de la orientación espacial, pero sin el soporte de hardware gráfico es difícil lograr un rendimiento interactivo. En esta aplicación, se utiliza una reciente técnica de iluminación de líneas que supera este inconveniente (Stalling *et al.*, 1997). Requiere un modelo de iluminación especial para líneas en $\Re^3$ (Banks, 1994) y basa la implementación del modelo en el mapeo de textura, característica que permite emplear un acelerador gráfico para los cálculos de sombreado.

En suma a una eficiente iluminación que facilita la percepción de la orientación espacial, mediante un método que permite la percepción de profundidad se proporciona una adecuada indicación de la ubicación espacial de la línea de flujo. Adicionalmente, su puede colorear la línea de flujo de acuerdo a un campo escalar, en este caso, es relevante visualizar el potencial eléctrico y la intensidad del vector densidad de corriente. Como las líneas de flujo de corriente se orientan del mayor a menor potencial, el mapeo del potencial en color es suficiente para permitir determinar el sentido de las líneas, cuando es ambiguo.

### 3.2 *Herramientas de visualización complementarias*

Se consideran algunas herramientas adicionales para la exploración interactiva de datos (Gallagher, 1995).

*Extractor de rebanadas*: Esta herramienta consiste en un plano que puede ser ubicado por el usuario en el espacio tridimensional. Sobre él, se mapea una cantidad escalar relevante codificada en color, la que se obtiene en la posición que ocupa el plano en el interior de la celda.

*Visualización del depósito de cobre*: Esta herramienta permite la visualización cualitativa del depósito de cobre en la superficie de los electrodos. Como este se relaciona directamente con la componente de la densidad de corriente normal a la superficie catódica, tal magnitud se mapea en color en las superficies catódicas de los electrodos. Para mayor flexibilidad de análisis, también se mapea la componente normal a las superficies anódicas.

*Cursor tridimensional*: Esta es una herramienta cuantitativa que permita conocer las magnitudes de los datos en el volumen de la celda. El cursor actúa como una sonda que entrega el valor de los datos correspondientes a su posición en la celda.

*Segmentación de datos*: Como las líneas de flujo se agrupan por electrodos, se puede implementar un simple y poderoso mecanismo poder eliminar problemas de oclusión de objetos de interés. Se basa en restringir la visualización a un conjunto reducido de datos seleccionados con el mouse.

## 4. IMPLEMENTACION Y RESULTADOS

Los fundamentos de la modelación y de visualización presentados en este artículo se usaron para desarrollar una herramienta computacional para la simulación y el análisis de celdas de EO, su GUI se muestra en la Fig. 2. Su desarrollo y los algoritmos numéricos empleados se detallan en (Mena, 2000). La aplicación se implemento en C++ utilizando la biblioteca gráfica OpenGL. Esta herramienta permite construir y analizar interactivamente cualquier geometría de celdas ocupando electrodos rectangulares, pudiendo ser unipolares y/o bipolares paralelos.

La concordancia de los resultados obtenidos al simular las celdas prototipos estudiadas en (Bittner, 1999), confirma que el modelo mejorado entrega resultados adecuados para el análisis de celdas de EO. En suma, el modelo mejorado es más preciso y eficiente.

En Fig. 2 a Fig. 4 se ilustran ejemplos del tipo de visualización de resultados que produce la aplicación. La iluminación de líneas de flujo y la sugestión de profundidad permiten apreciar adecuadamente la estructura del campo de densidad de corriente. La Fig. 2 ilustra como el color de las líneas de flujo, codificando el potencial eléctrico, se puede usar para indicar el sentido de las líneas de flujo cuando resulta ambiguo (en este caso, desde amarillo a rojo).

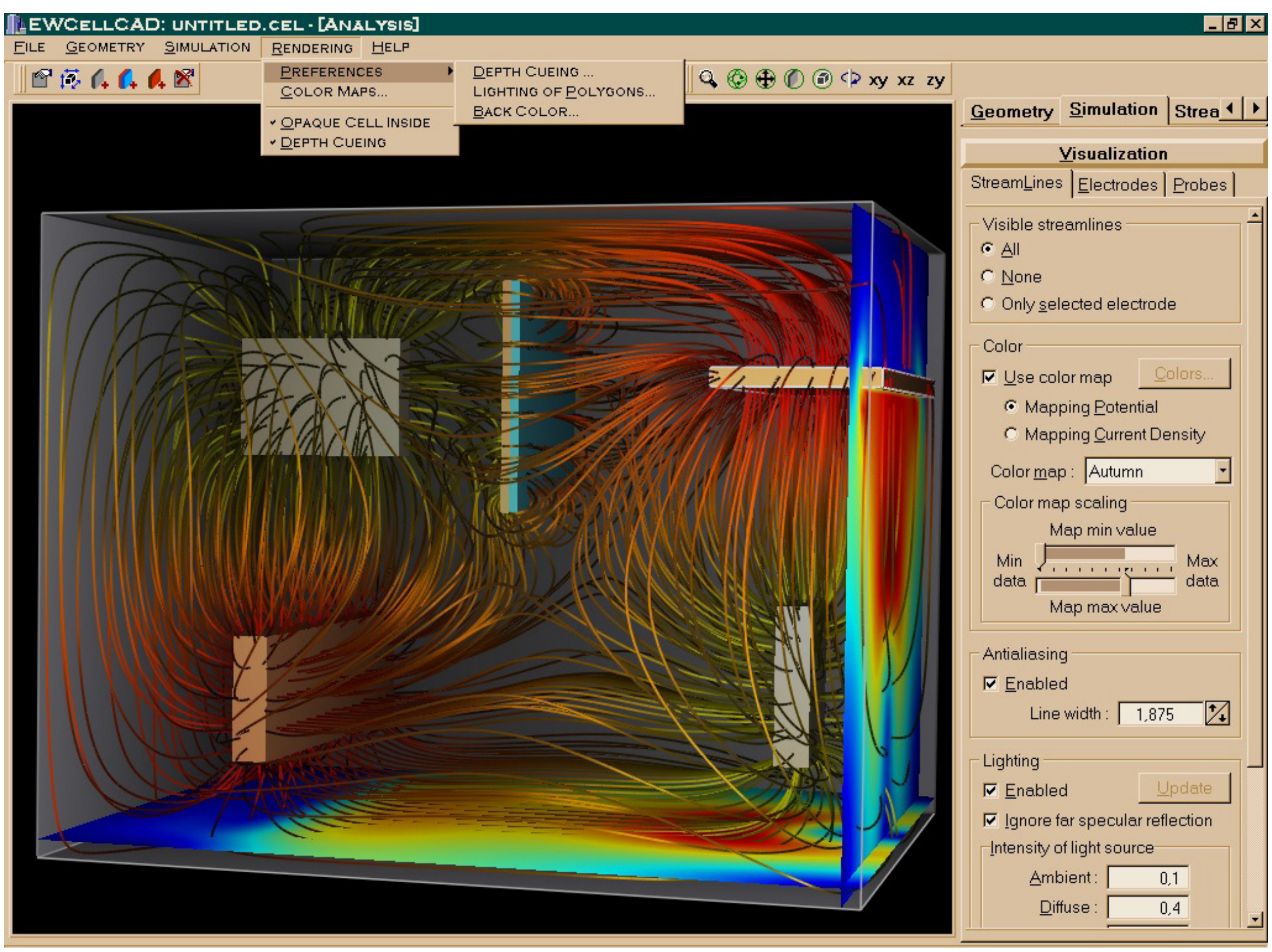

Fig. 2. Interfaz Gráfica de Usuario (GUI) de la aplicación desarrollada para el análisis de celdas de EO.

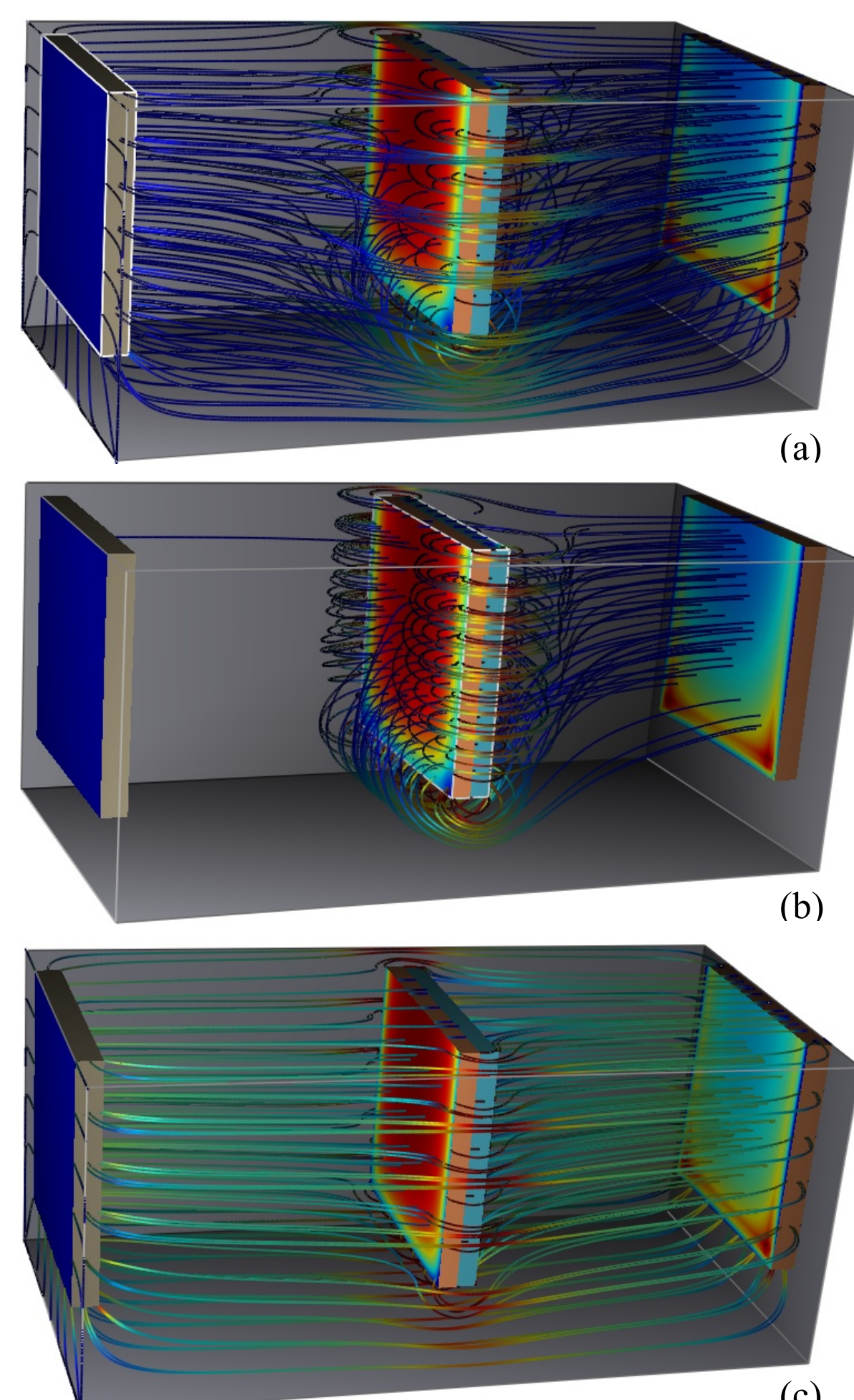

Fig. 3. Celda de EO con un electrodo bipolar. (a) - (b) Electrodo bipolar débilmente polarizado. (b) Líneas de flujo generadas en electrodo bipolar muestran que no es atravesado por corriente que sale del ánodo. (c) Electrodo bipolar polarizado.

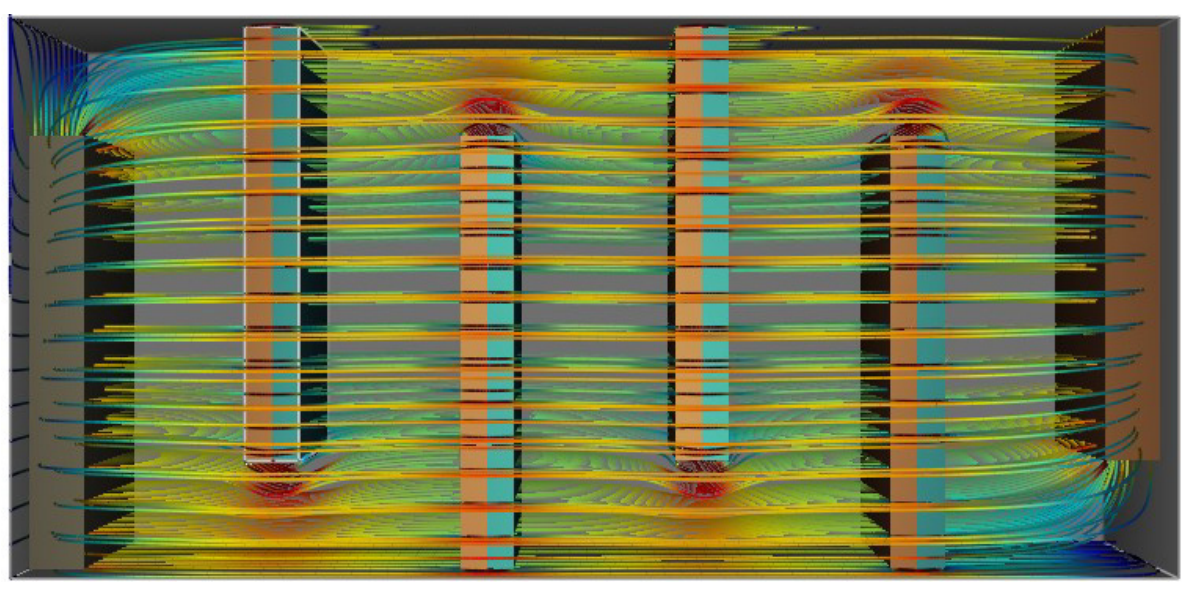

Fig. 4. Celda de EO asimétrica con 4 electrodos bipolares. Vista ortogonal inferior.

## 5. CONCLUSION

Para estudiar diferentes estructuras geométricas, de modo de optimizar el proceso de EO de cobre usando electrodos bipolares flotantes, es importante disponer de una herramienta de simulación y análisis, basada en un adecuado modelo del proceso. En este artículo se ha presentado los fundamentos con los cuales se implementó una aplicación computacional orientada al logro de dicha herramienta.

Con objeto de obtener resultados más precisos, se contribuyó a mejorar el modelo eléctrico disponible. En particular, el modelo de electrodo bipolar, el que se obtuvo a partir de una formulación analítica para calcular el potencial metálico en electrodos flotantes.

La aplicación desarrollada proporciona una interfaz gráfica de usuario que facilita el diseño geométrico de las celdas, permite simular la celda con control del error numérico y proporciona herramientas de visualización adecuadas para analizar los datos generados por el nuevo modelo eléctrico. En particular, se aplicó una conveniente técnica para la visualización interactiva de campos vectoriales.

## REFERENCIAS